\documentclass[%
 aip,
 jmp,%
 amsmath,amssymb,
 reprint,%
]{revtex4}

\draft 

\begin{document}


\title{The Oscillator Strengths of $\mbox{H}^+_2$, $1s\sigma_{g}-2p\sigma_{u}$, $1s\sigma_{g}-2p\pi_{u}$} 

\author{Ts.~Tsogbayar}
 \altaffiliation[Permanent Address: ]{Institute of Physics and Technology, Mongolian Academy of Sciences,
 Peace Ave 54B, 210651, Ulaanbaatar, Mongolia}
\affiliation{Department of Physics and Astronomy, York University,
4700 Keele Street, Toronto, Ontario, M3J 1P3 Canada}

\author{Ts.~Banzragch}%
\affiliation{
 Department of Theoretical and Experimental Physics, School of Physics and Electronics,
 National University of Mongolia,
 Ikh surguuliin gudamj 1, Ulaanbaatar, Mongolia
}%

\date{\today}

\begin{abstract}
The oscillator strengths of the $\mbox{H}^+_2$ molecular ion,
$1s\sigma_{g}-2p\sigma_{u}$, $1s\sigma_{g}-2p\pi_{u}$ are
calculated within the Born-Oppenheimer approximation. The
variational expansion with randomly chosen exponents has been used
for numerical studies. The oscillator strengths obtained for the
transitions $1s\sigma_{g}-2p\sigma_{u}$, $1s\sigma_{g}-2p\pi_{u}$
of $\mbox{H}^+_2$ are accurate up to ten significant digits.
Results are given for the internuclear distances between $0.10$
and $20.0$\,a.u.
\end{abstract}

\pacs{}

\maketitle 

\section{Introduction}
  Many theoretical calculations on the oscillator strengths of the $\mbox{H}^+_2$ molecular ion have been
 performed since the publication of the first \cite{Mull} of Mulliken's series
 of papers on the subject. Bates and his co-workers had done a great deal of work on this molecular system, with
 special emphasis on the electronic transition probabilities for fixed internuclear separation $R$ for
 a large number of electronic transitions \cite{Bates, Bates2,
 Bates3, Bates4, Dalg}. The problem was formulated in terms of Jaffe's solution
 \cite{Jaffe} and led to certain integrals in the expression for the transition
 probabilities which required numerical evaluation. Moreover,
 using the Hylleraas solution \cite{Hyll}  the matrix elements of the electric dipole moment for transition
 between bound electronic states of the $\mbox{H}^+_2$ molecular
 ion for any fixed internuclear separation $R$ have been obtained
 in analytical form \cite{Her}. In the last years, with the aid of the new types of
 variational expansions based on the randomly chosen exponents for the Coulomb three--body problem \cite{Korobov1, Bai}, within
the Born-Oppenheimer approximation, very accurate non-relativistic
 energies and the improved relativistic corrections of $m\alpha^{4}$ and $m\alpha^6$
 orders for the ground state of the ion, as well the more accurate static dipole polarizability for the $1s\sigma_{g}$
 electronic state of $\mbox{H}^+_2$ molecular ion have been
 obtained in \cite{Tsogbayar1,Korobov2,Tsogbayar2}.

 In this work our goal is to obtain more accurate values for the oscillator
 strengths of the $1s\sigma_{g}-2p\sigma_{u}$,
$1s\sigma_{g}-2p\pi_{u}$ transitions of $\mbox{H}^+_2$ molecular
ion within the Born-Oppernheimer approximation than those in
\cite{Bates} and \cite{Ram}.

We want to show that the use of the variational expansion
suggested in this work allows an analytical evaluation of the
matrix elements and can provide us with very accurate data. In
future studies it can be used to obtain more accurate values of
moments functions $S_{k}(R)$ as a function of internuclear
separation $R$ for the $1s\sigma_{g}$ electronic state of the
$\mbox{H}^+_2$ molecular ion in the Born-Oppenheimer
approximation.

\section{Theory}

In what follows we consider the Coulomb two--center problem with
the nonrelativistic Hamiltonian
\begin{equation}\label{nonrel}
H = \frac{\mathbf{p}^2}{2m}+V, \qquad V =
-\frac{Z_1}{r_1}-\frac{Z_2}{r_2}\,.
\end{equation}
where $r_1$ and $r_2$ are the distances from an electron to nuclei
1 and 2, respectively.

The oscillator strength $f_{0n}$ (details can be found in
\cite{Mull, Bates}) can be expressed as
\begin{equation}\label{osc}
 f_{0n} = \frac{2}{3} G |Q|^2(E_{n} - E_{0})
\end{equation}
where $G$ is the orbital degeneracy factor, and $G=1$ for $\sigma$
states, $G=2$ for $\pi$ states. Here $Q$ is the transition moments
and can be written in the form
\begin{equation}\label{trans_m}
  Q = \left\langle 0| r|n \right\rangle ,
\end{equation}
where $r$ is a distance of the electron from the center along the
internuclear axis:
$$
 r = \frac{1}{2 R}\sqrt{2
 r^{2}_{1}r^{2}_{2}+2r^{2}_{1}R^{2}+2r^{2}_{2}R^{2}-r^{4}_{1}-r^{4}_{2}-R^{4}}
 \,.
$$

\section{Variational approximation and numerical results}

In order to get a precise solution for the Schr\"{o}dinger
equation
\begin{equation}\label{SE}
  \Big[\frac{\mathbf{p}^2}{2 m} + V \Big]\Psi_{0}(\mathbf{r}) =
  E_{0} \Psi_{0}(\mathbf{r})
\end{equation}
we use the variational approach. A variational wave function for
the electronic states of the $\mbox{H}^+_2$ should be symmetrized
and is constructed as follows:
\begin{equation}\label{wf}
   \Psi(\mathbf{r_{1},r_{2}}) = e^{i m\varphi } r^{|m|} \sum^{\infty}_{i=1}
 C_{i}(e^{-\alpha_{i} r_{1} - \beta_{i} r_{2}}\pm
       e^{-\beta_{i} r_{1} - \alpha_{i} r_{2}} ),
\end{equation}
where $(+)$ is used for the \textit{gerade} electronic state and
$(-)$ is for the \textit{ungerade} electronic state, respectively.
Parameters $\alpha_{i}$ and $\beta_{i}$ are generated in
quasi-random manner~\cite{Korobov1, Bai}
\begin{equation}\label{1.5}
  \alpha_{i}=\lfloor\frac{1}{2}i(i+1)\sqrt{p_{\alpha}}\rfloor(A_{2}-A_{1})
  + A_{1}
\end{equation}
$\lfloor x\rfloor$ designates the fractional part of $x$,
$p_{\alpha}$ is a prime number, an interval $[A_{1},A_{2}]$ is a
real variational interval, which has to be optimized. Parameters
$\beta_{i}$ are obtained in a similar way.

Numerical evaluation of the matrix elements for operators in
(\ref{osc}) and (\ref{trans_m}) is expounded in the Appendix.

\begin{table}[h]
\caption{Data on the $1s\sigma_{g}-2p\sigma_{u}$,
$1s\sigma_{g}-2p\pi_{u}$ transitions of $H^{+}_{2}$}
\begin{center}
\begin{tabular}{c@{\hspace{3mm}}c@{\hspace{4mm}}c@{\hspace{4mm}}
   c@{\hspace{4mm}}c@{\hspace{4mm}}c@{\hspace{4mm}}c}
\hline \hline & \multicolumn{3}{c}{Bates \cite{Bates}} & &
{Ramaker and Peek \cite{Ram}}
\\ \cline{3-3}
\cline{6-6}
$R$ & $E_{1} - E_{0}$  & $f_{01}$ & $f_{01}$ & $E_{2}-E_{0}$ & $f_{02}$ & $f_{02}$\\
\hline
 0.10 & 0.147\,757\,4626(1) &       & 0.141\,946\,1197 & 0.147\,857\,3935(1) &       & 0.281\,958\,4589\\
 0.20 & 0.142\,594\,2876(1) & 0.150 & 0.149\,966\,1487 & 0.142\,993\,1822(1) &       & 0.292\,176\,8241\\
 0.30 & 0.136\,065\,8191(1) &       & 0.161\,227\,2533 & 0.136\,960\,1755(1) &       & 0.305\,026\,4125\\
 0.40 & 0.128\,996\,9831(1) & 0.175 & 0.174\,800\,4051 & 0.130\,578\,7951(1) &       & 0.318\,871\,6774\\
 0.50 & 0.121\,810\,2535(1) &       & 0.190\,006\,5537 & 0.124\,264\,4532(1) &       & 0.332\,827\,2240\\
 0.60 & 0.114\,717\,4343(1) & 0.207 & 0.206\,235\,8257 & 0.118\,218\,5058(1) &       & 0.346\,408\,7672\\
 0.70 & 0.107\,822\,0761(1) &       & 0.222\,877\,7173 & 0.112\,529\,8372(1) &       & 0.359\,356\,2504\\
 0.80 & 0.101\,173\,4174(1) & 0.240 & 0.239\,317\,2609 & 0.107\,228\,2899(1) &       & 0.371\,539\,1200\\
 0.90 & 0.947\,946\,4449    &       & 0.254\,970\,6910 & 0.102\,313\,0356(1) &       & 0.382\,902\,9462\\
 1.00 & 0.886\,972\,6883    & 0.269 & 0.269\,336\,9702 & 0.977\,678\,3571    & 0.393 & 0.393\,438\,1322\\
 1.30 & 0.721\,475\,3157    &       & 0.301\,731\,0394 & 0.861\,082\,3079    &       & 0.420\,300\,6176\\
 1.50 & 0.625\,821\,6700    & 0.314 & 0.313\,859\,3007 & 0.797\,263\,3391    & 0.435 & 0.434\,630\,6848\\
 1.80 & 0.503\,578\,1216    & 0.321 & 0.320\,620\,5783 & 0.717\,917\,7718    & 0.451 & 0.451\,551\,8442\\
 2.00 & 0.435\,099\,8223    & 0.319 & 0.319\,763\,3919 & 0.673\,862\,3946    & 0.460 & 0.460\,186\,9855\\
 2.30 & 0.349\,292\,2274    &       & 0.313\,481\,1372 & 0.618\,286\,6408    &       & 0.469\,722\,4456\\
 2.50 & 0.301\,751\,4622    & 0.307 & 0.307\,111\,6776 & 0.587\,007\,7894    & 0.474 & 0.474\,048\,0566\\
 2.80 & 0.242\,384\,3555    & 0.297 & 0.295\,449\,1732 & 0.547\,111\,4888    & 0.478 & 0.477\,811\,5429\\
 3.00 & 0.209\,477\,8640    & 0.289 & 0.286\,648\,3110 & 0.524\,452\,3451    & 0.479 & 0.478\,648\,2112 \\
 3.50 & 0.145\,359\,8123    & 0.264 & 0.262\,040\,2032 & 0.478\,772\,9801    & 0.475 & 0.475\,449\,5457 \\
 4.00 & 0.100\,534\,2444    & 0.238 & 0.234\,600\,2369 & 0.445\,260\,2396    & 0.465 & 0.465\,523\,7278 \\
 4.50 & 0.691\,110\,5645($-$1)  & 0.295 & 0.205\,243\,3086 & 0.420\,759\,4992 & 0.450 & 0.449\,971\,8773 \\
 5.00 & 0.471\,286\,8194($-$1)  & 0.175 & 0.175\,094\,9934 & 0.403\,035\,4828 & 0.430 & 0.430\,144\,3360 \\
 5.50 & 0.318\,483\,6379($-$1)  & 0.144 & 0.145\,512\,8632 & 0.390\,423\,8948 & 0.408 & 0.407\,637\,0403 \\
 6.00 & 0.213\,251\,5610($-$1)  & 0.116 & 0.117\,844\,6702 & 0.381\,644\,1528 & 0.384 & 0.384\,106\,9351 \\
 6.50 & 0.141\,551\,5325($-$1)  & 0.090 & 0.931\,480\,9973($-$1) & 0.375\,699\,5692 & 0.361 & 0.361\,026\,7789 \\
 7.00 & 0.932\,229\,1695($-$2)  & 0.069 & 0.720\,267\,2925($-$1) & 0.371\,819\,5628 & 0.339 & 0.339\,507\,0520 \\
 7.50 & 0.609\,732\,1052($-$2)  & 0.090 & 0.546\,252\,3165($-$1) & 0.369\,418\,4230 & 0.320 & 0.320\,242\,4604 \\
 8.00 & 0.396\,437\,2998($-$2)  & 0.039 & 0.407\,352\,9645($-$1) & 0.368\,059\,7573 & 0.304 & 0.303\,561\,4626 \\
 8.50 & 0.256\,445\,4385($-$2)  & 0.028 & 0.299\,374\,4120($-$1) & 0.367\,424\,2386 & 0.289 & 0.289\,523\,5398 \\
 9.00 & 0.165\,162\,3429($-$2)  & 0.020 & 0.217\,256\,4226($-$1) & 0.367\,281\,4212 & 0.278 & 0.278\,018\,0087 \\
 9.50 & 0.105\,969\,2322($-$2)  & 0.016 & 0.155\,938\,7742($-$1) & 0.367\,466\,5422 & 0.269 & 0.268\,841\,6239 \\
10.00 & 0.677\,660\,3412($-$3)  & 0.011 & 0.110\,852\,7572($-$1) & 0.367\,862\,4388 & 0.262 & 0.261\,750\,5281 \\
11.00 & 0.274\,807\,5098($-$3)  &       & 0.546\,544\,7478($-$2) & 0.368\,978\,5760 &       & 0.252\,815\,1953 \\
12.00 & 0.110\,412\,3031($-$3)  &       & 0.262\,186\,7652($-$2) & 0.370\,217\,2763 &       & 0.249\,293\,6287 \\
13.00 & 0.440\,274\,9253($-$4)  &       & 0.122\,982\,9099($-$2) & 0.371\,377\,5187 &       & 0.249\,503\,7766 \\
14.00 & 0.174\,459\,4699($-$4)  &       & 0.566\,128\,7073($-$3) & 0.372\,370\,6777 &       & 0.252\,027\,6563 \\
15.00 & 0.687\,612\,7753($-$5)  &       & 0.256\,469\,7079($-$3) & 0.373\,170\,2448 &       & 0.255\,725\,8261 \\
16.00 & 0.269\,771\,2233($-$5)  &       & 0.114\,593\,6419($-$3) & 0.373\,783\,9614 &       & 0.259\,757\,3081 \\
17.00 & 0.105\,416\,2788($-$5)  &       & 0.505\,885\,9722($-$4) & 0.374\,236\,6462 &       & 0.263\,577\,2171 \\
18.00 & 0.410\,477\,9229($-$6)  &       & 0.220\,971\,2567($-$4) & 0.374\,559\,1664 &       & 0.266\,896\,4478 \\
19.00 & 0.159\,336\,3912($-$6)  &       & 0.956\,149\,8035($-$5) & 0.374\,781\,7742 &       & 0.269\,614\,3011 \\
20.00 & 0.616\,776\,4338($-$7)  &       & 0.410\,256\,5319($-$5) & 0.374\,930\,7098 &       & 0.271\,746\,9205 \\
\hline \hline
\end{tabular}
\end{center}
\end{table}

\begin{table}[h]
\caption{Convergence of the $1s\sigma_{g}-2p\sigma_{u}$ and
$1s\sigma_{g}-2p\pi_{u}$ transitions with an increase in the
basic-set size of $\Psi$}
\begin{center}
\begin{tabular}{c@{\hspace{3mm}}c@{\hspace{4mm}}c@{\hspace{4mm}}c@{\hspace{4mm}}}
\hline\hline
$R (a.u.)$ & $ N $ & $f_{01}$  & $f_{02}$   \\
\hline
0.10 & 110\, 65 & 0.141\,946\,119\,692\,60 & 0.281\,958\,458\,927\,66\\
     & 125\, 80 & 0.141\,946\,119\,692\,60 & 0.281\,958\,458\,927\,67\\
     & 140\, 95 & 0.141\,946\,119\,692\,60 & 0.281\,958\,458\,927\,67\\
2.00 & 90\, 68 & 0.319\,763\,391\,895\,63 & 0.460\,186\,985\,489\,56\\
     & 120\, 78 & 0.319\,763\,391\,895\,63 & 0.460\,186\,985\,489\,56\\
     & 140\, 98 & 0.319\,763\,391\,895\,63 & 0.460\,186\,985\,489\,56\\
20.0 & 90\, 105 & 0.410\,256\,531\,872\,29(-5) & 0.271\,746\,920\,513\,63\\
     & 110\, 125 & 0.410\,256\,531\,861\,97(-5) & 0.271\,746\,920\,515\,47\\
     & 130\, 160 & 0.410\,256\,531\,861\,96(-5) & 0.271\,746\,920\,515\,47\\
\hline\hline
\end{tabular}
\end{center}
\end{table}

\begin{table}[h]
\caption{Comparison with earlier calculations at a bond length
$R=2.0\,a.u.$}
\begin{center}
\begin{tabular}{@{}c@{\hspace{6mm}}l@{\hspace{6mm}}l@{\hspace{6mm}}l@{}}
\hline\hline
 & \multicolumn{1}{c}{$f_{01}$~~~~~~~~~~~~~~~~~} & \multicolumn{1}{c}{$f_{02}$~~~~~~~~~~~~~}
 \\
\hline
Mulliken \cite{Mull}
& $0.31$ & $ $  \\
Dalgarno and Poots \cite{Dalg}
& $0.319$ & $ $ \\
Bates \textit{et al} \cite{Bates, Bates3}
& $0.319$ & $0.460$ \\
Ramaker and Peek \cite{Ram}
& $0.319$ & $0.460$ \\
This work
& $0.319\,763\,3919$ & $0.460\,186\,9855$ \\
\hline\hline
\end{tabular}
\end{center}
\end{table}

In Tables I the energy differences and and oscillator strengths as
a function of internuclear separation $R$ for
$1s\sigma_{g}-2p\sigma_{u}$ transitions ($E_{1}-E_{0}$ and
$f_{01}$) and $1s\sigma_{g}-2p\pi_{u}$ transitions ($E_{2}-E_{0}$
and $f_{02}$) and comparisons with the previous results are
presented. The estimated accuracy of obtained values for the
transitions is ten significant digits. In Table II the convergence
of the $1s\sigma_{g}-2p\sigma_{u}$ and $1s\sigma_{g}-2p\pi_{u}$
transitions with an increase in the expansion length of the $\Psi$
(equation (\ref{wf})) is demonstrated for some values of
internuclear distance $R$.

In order to get accurate results we use three sets of basic
functions of the type (\ref{wf}) (in a spirit of
Ref.\cite{Korobov1}) for small values of internuclear distance
$R$, two sets for intermediate and large values of $R$,
respectively. Total number of the basic function varies from
$N=90$ to $N=180$. In Table II in the second and third columns the
numbers of the basis functions employed (equation (\ref{wf})) for
the electronic states $2p\sigma_{u}$ and $2p\pi_{u}$ of
$\mbox{H}^+_2$ are presented, respectively. In our calculations
arithmetics of sextuple precision (about 48 decimal digits)
implemented as a FORTRAN90 module has been used.

In all tables the factor $x$ in the brackets means $10^{x}$.
Atomic units are used throughout.

\section{Conclusion}
The oscillator strengths of the $\mbox{H}^+_2$ molecular ion,
$\mbox{H}^+_2$, $1s\sigma_{g}-2p\sigma_{u}$,
$1s\sigma_{g}-2p\pi_{u}$ have been calculated accurately within
the Born-Oppenheimer approximation. The variational expansion with
randomly chosen exponents has been used for numerical studies.
This type of expansion allows us to use few number of basic
functions, and even the non-relativistic energy for the
$1s\sigma_{g}$ electronic state of the ion can be obtained to
$10^{-20}$a.u. accuracy when the number of basis functions is less
than $N=100$ \cite{Tsogbayar2}.

Previous calculations for the $1s\sigma_{g}-2p\sigma_{u}$
transitions of the ion for a fairly-restricted internuclear
distance $R$ are by Mulliken \cite{Mull}, Bates \cite{Bates}, and
Dalgarno and Poots \cite{Dalg}. Mulliken \cite{Mull} used the LCAO
approximation and Bates \cite{Bates} employed the exact two-center
wave function for the numerical integral evaluation. Dalgarno and
Poots \cite{Dalg} employed the variational method in the
calculation. The calculations for the $1s\sigma_{g}-2p\pi_{u}$
transition of $\mbox{H}^+_2$ are by Bates, Darling, Hawe and
Stewart \cite{Bates3}, Herman and Wallis \cite{Her} and Ramaker
and Peek \cite{Ram}. Bates \textit{et al} \cite{Bates3} used also
the exact two-center wave function for the numerical integration.
Herman and Wallis \cite{Her} obtained the matrix elements for the
transitions between bound electronic states of the ion for any
fixed internuclear distance $R$ in an analytical form using the
Hylleraas solution for the wave functions. Ramaker and Peek
\cite{Ram} performed the calculations for the dipole strengths of
the $\mbox{H}^+_2$ molecular ion as functions of wide range of
internuclear separation $R$ for all transitions the states which
correlate to a proton-plus-hydrogen-atom and have principal
quantum number $n=1, 2$ or $3$, using the exact electronic
two-center wave function.

In Table III, we compare our results with earlier ones, which
demonstrate the superiority of the newly obtained results. The
results obtained for the transitions $1s\sigma_{g}-2p\sigma_{u}$,
$1s\sigma_{g}-2p\pi_{u}$ of $\mbox{H}^+_2$ are accurate up to ten
significant digits, that is, six digits more than in \cite{Bates,
Ram}. With the help of the results obtained in this
\textit{Letter}, one can recalculate the mean excitation energy
for the $1s\sigma_{g}$ electronic state of the $\mbox{H}^+_2$
molecular ion employing the functional form of the moment
functions $S_{k}(R)$ \cite{Gar,Gers,Bish}. This work is in
progress now.

\appendix
\section*{Appendix: Analytical evaluation of the matrix elements}
\renewcommand{\theequation}{A-\arabic{equation}}
\setcounter{equation}{0}

The calculation of the matrix elements is reduced to evaluation of
integrals of the type
\begin{equation}
 \Gamma_{lm}(\alpha,\beta) =
   \int r^{l-1}_{1}r^{m-1}_{2}e^{-\alpha r_{1}-\beta r_{2}}
                                            d^{3}\mathbf{r}.
\end{equation}
Integers $(l, m)$ are, in general, non-negative, but in case of
singular matrix elements one of the indices can be negative.

The function $\Gamma_{00}$ can be easily obtained
\begin{equation}\label{A.2}
\Gamma_{00}(\alpha,\beta,R) = \frac{4\pi}{R}\,
    \frac{e^{-\beta R} - e^{-\alpha R}}{\alpha^{2} - \beta^{2}},
\end{equation}
where $R$ is the distance between nuclei, then
$\Gamma_{lm}(\alpha,\beta;R)$ for non-negative $(l, m)$ may be
generated from (\ref{A.2}) by means of relation
\begin{equation}\label{A.3}
\Gamma_{lm}(\alpha,\beta;R) =
  \left( - \frac{\partial}{\partial \alpha}\right)^{l}
  \left( - \frac{\partial}{\partial \beta}\right)^{m}
  \Gamma_{00}(\alpha,\beta,R).
\end{equation}

Integral $\Gamma_{-1,0}(\alpha,\beta;R)$ is expressed by
\begin{equation}\label{A.4}
\begin{array}{@{}l}
\displaystyle \Gamma_{-1,0}(\alpha,\beta;R) =
   \frac{2\pi}{R\beta}
      \Bigl\{
         e^{\beta R} \mbox{E}_{1}(R(\alpha+\beta))
         +e^{-\beta R}\ln R(\alpha +\beta)
\\[3mm]\hspace{25mm}
         -e^{\beta R}\bigl[\mbox{E}_{1}(R(\alpha -\beta))
         +\ln R(\alpha -\beta)\bigr]
      \Bigr\}.
\end{array}
\end{equation}
Worthy to note that a function in square brackets is analytic when
argument is zero. Integrals $\Gamma_{-1,m}$ are generated from
$\Gamma_{-1,0}$ similar to (\ref{A.3}):
\begin{equation}\label{A.5}
\Gamma_{-1,m}(\alpha,\beta;R) =
  \left( - \frac{\partial}{\partial \beta}\right)^{m}
  \Gamma_{-1,0}(\alpha,\beta,R).
\end{equation}

Function $\mbox{E}_1(z)$ encountered in (\ref{A.4}) is the
exponential integral function \cite{Abr}:
\[
 \mbox{E}_{1}(z) = \Gamma(0,z) = \int^{\infty}_{z} t^{-1}e^{-t}dt.
\]

\bibliography{your-bib-file}

\end{document}